\documentclass{emulateapj}
\usepackage{graphicx}
\usepackage{natbib}
\usepackage{color}
\usepackage{mathtools}
\bibpunct{(}{)}{;}{a}{}{,}

\shorttitle{BLR formation mechanism in AGN}
\shortauthors{Czerny et al.}
\begin{document}


\title{Test of the formation mechanism of the Broad Line Region in Active Galactic Nuclei}


\author{Bozena Czerny\altaffilmark{1,2}, Pu Du\altaffilmark{2}, Jian-Min Wang\altaffilmark{2} and Vladimir Karas\altaffilmark{3}}
\email{bcz@cft.edu.pl}

\altaffiltext{1}{Center for Theoretical Physics,
    Polish Academy of Sciences, Al. Lotnik\' ow 32/46,
02-668 Warsaw, Poland} 

\altaffiltext{2}{Key Laboratory for Particle Astrophysics, Institute of High Energy Physics, Chinese Academy of Sciences, 19B Yuquan Road, Beijing 100049, China}

\altaffiltext{3}{Astronomical Institute, Academy of Sciences, Bocni II 1401, CZ-141 00 Prague, Czech Republic} 

\begin{abstract}
The origin of the Broad Line Region (BLR) in active galaxies remains unknown. It seems to be related to the underlying accretion disk but an efficient mechanism is required to rise the material from the disk surface without giving too strong signatures of the outflow  in the case of the low ionization lines. We discuss in detail two proposed mechanisms: (i) radiation pressure acting on dust in the disk atmosphere creating a failed wind (ii) the gravitational instability of the underlying disk. We compare the predicted location of the inner radius of the BLR in those two scenarios with the observed position obtained from the reverberation studies of several active galaxies. The failed dusty outflow model well represents the observational data while the predictions of the self-gravitational instability are not consistent with observations. The issue remains why actually we do not see any imprints of the underlying disk instability in the BLR properties.
\end{abstract}

%% Keywords should appear after the \end{abstract} 
%%command. The uncommented
%% example has been keyed in ApJ style. See the 
%%instructions to authors
%% for the journal to which you are submitting 
%%your paper to determine


\section{Introduction}

Broad Emission Lines in the IR/optical/UV band are the most characteristic features of radio-quiet Active Galactic Nuclei (AGN). They originate from the Broad Line Region (BLR) typically located a fraction of a parsec from the central black hole. The general view is that the emission comes partially from the irradiated surface of the accretion disk surrounding the supermassive black hole and partially, or mostly, from the clouds located above the disk.

Apart from being of interest per se, the BLR observations and modeling is crucial as it currently provides the best way to measure the black hole masses in AGN (see Peterson 2014, and the references therein), and the distances to AGN, which may be efficiently used in the future for cosmological applications (Watson et al. 2011; Czerny et al. 2013).

The emitted lines represent broad range of the physical parameters, and the Doppler effects indicate a range of velocities so the region is certainly stratified. It is frequently divided into two main components: Low Ionization Line (LIL) Region and High Ionization Line (HIL) Region (Collin-Souffrin et al. 1988). HIL  (e.g. He II and CIV lines) are frequently shifted with respect to the systemic redshift and likely come from some outflowing wind (e.g. Kollatschny 2003). LIL show much less or no shift, and they are frequently considered to originate much closer to the accretion disk surface.  They also originate at larger distances from a black hole, as implied by their line profiles (e.g. Kollatschny 2003) and time delay with respect to the continuum (e.g. Wanders et al. 1997; Wandel et al. 1999; Trevese et al. 2014). The properties of the LIL part of the BLR is  thus especially puzzling, since (i) their covering factor is quite large so the corresponding clouds should be located high above the disk (ii) they do not come from the wind (little Doppler shift) (iii) they must be subject of radiation pressure (iv) their densities are high.  Wind model predicts appreciable blueshift excess in the line profiles (see Waters et al. and the references therein) while Narrow Line Seyfert 1 galaxies show symmetric lines, well modeled with a single Lorentzian shape (e.g. Laor et al. 1987; Cracco et al. 2016). Some Broad Line Seyfert 1 galaxies and quasars show an additional, separate, blueshifted component (see e.g. Sulentic et al. 2015) which is related to the wind).  The H$\beta$ line, used for the black hole mass determination in nearby AGN belongs to LIL class. Also most of the available reverberation measurements of the BLR size are based on H$\beta$ (see the compilation by Misty Bentz\footnote{http://www.astro.gsu.edu/AGNmass/}. Therefore, we may say we concentrate on the properties of the LIL part of the BLR in the current model, although our study is of a more general character, focusing on constraining the plausible BLR formation mechanism.

The overall properties of the BLR are successfully described by the Locally Optimized Cloud (LOC) model (Baldwin et al. 1995). However, in this model a whole family of clouds is set through a parametric description, with overall normalizations and power law dependencies of the cloud density and radial location being arbitrary parameters, to be fitted against the data.

However, the existence of the clouds so high above the disk requires some explanation, and two specific models were developed which give the quantitative predictions where the inner radius of the BLR should be located in a given AGN.

The first model is failed radiatively accelerated dusty outflow (FRADO) model, where the radiation pressure acts more strongly in the disk surface layers when the effective temperature in the disk falls below the dust sublimation temperature (Czerny \& Hryniewicz 2011, Czerny et al. 2015). This alleviates the disk material up, but higher above the disk the material is strongly irradiated by the central region, dust evaporates and the material falls down.

The second model is based on the concept that the accretion disk in an AGN become self-gravitating at some distance from the black hole. There the instability develops which might lead to vigorous star formation (Collin \& Zahn 1999, 2008;  Levin \& Beloborodov 2003; Wang et al. 2012), and this increased turbulence rises up the material from the disk. We will refer to this model as self-gravitating model.

Both models give specific predictions for the location of the inner radius of  BLR within the frame of the $\alpha-$disk model, when the black hole mass, accretion rate, and the viscosity parameter $\alpha$ are known. This can be compared with observational results for the BLR size from reverberation measurement (RM) campaigns. Thus the aim of our paper is to confront those two models with observational data to see which one better represents the observed trend.

\section{Models}

We consider two likely models of the formation of the BLR: FRADO model, based on the idea of the key role of the dust in the disk atmosphere, and the self-gravity models, based on the idea of strong turbulence in the disk caused by self-gravity. Both mechanisms could be responsible for the efficient rise of the disk material above the equatorial plane and, in consequence, for its exposure to the central source and efficient line formation. The radial range where these mechanisms start to operate can be calculated from the properties of accretion disks, and these radii can be compared with the time delay measurements. 
 
Since we are interested in the properties of accretion disk at distances higher than $\sim 100$ Schwarzschild radii ($R_{Schw}$) the relativistic effects close to the black hole horizon can be neglected. We thus use the basis equations of the stationary disk vertical structure as formulated in Shakura \& Sunyaev (1973), with the appropriate modifications. 

In the case of FRADO model, the location of the inner radius of the BLR is simply set by the global parameters: black hole mass and accretion rate, and by the assumption about the dust sublimation temperature. In this case we do not consider the self-gravity effects in the disk, and the results do not require the knowledge of the viscosity or the vertical disk structure.

In the case of the self-gravity model, computations of the disk structure, with appropriate modifications due to self-gravity, are required. 

\subsection{FRADO model}

The predictions of the FRADO model can be done with the standard Shakura-Sunyaev disk model without any modifications. The disk model is specified by the value of the black hole mass, $M$, and accretion rate, $\dot M$.  The locally emitted flux, $F$, at a radius $r$, integrated over frequency, is then given by the conservation laws of the mass and angular momentum
\begin{equation}
F = {3 G M \dot M \over 8 \pi r^3} \left(1 - \sqrt{3 R_{Schw}\over r}\right),
\end{equation} 
and the local effective temperature of the disk is
\begin{equation}
F = \sigma_B T_{eff}^4 
\end{equation}
where $\sigma_B$ is the Stephan-Boltzman constant. We further assume that the disk radiates as a black body, since color-correction to the effective temperature is only important in the innermost parts of the disk (e.g. Czerny et al. 2011; Done et al. 2012). 

In the case of FRADO model the inner radius of the BLR is set by the dust sublimation temperature, $T_{dust}$. 
This temperature in general is not known very precisely, and it depends on the chemical composition of the grain, and the size, as well as on the local pressure,  and it is not well established.  In general, silicates evaporate in the temperature as low as $\sim 800$ K, while amorphous grain, and particularly graphite, can exists at much higher temperatures. The maximum temperature of the dust in the solar system spans the range 1370 - 1770 K (Posch et al. 2007), and the measured dust temperature in the case of the active galaxy NGC 4151 varied with the nuclear flux, reaching up to 1500 K (Schnuelle et al. 2013).  In Czerny \& Hryniewicz (2011) we used 1000 K as a reference, but Galianni \& Horne (2013) in his study of NGC 5548 advocated 1670 K, with the difference due to different assumptions about the geometry. In the model below, $T_{dust}$ is assumed to be a free parameter of the model. 

Thus in FRADO model the inner radius of the BLR is set by the condition
\begin{equation}
T_{eff}(r) = T_{dust},
\end{equation}
and can be calculated for an assumed  $T_{dust}$ as a function of the black hole mass and accretion rate.  Here we assume that the dust and gas are thermally coupled which is justified in the upper disk layers and in the BLR clouds. The densities there are of order $10^9$ cm$^{-3}$ or higher (see e.g. Adhikari et al. 2016).

In observations, the delay is measured against the monochromatic flux (at  5100 \AA~ in the case of H$\beta$ monitoring).  This flux can be also calculated from the model from a given black hole mass and accretion rate. In the long wavelength limit, where the frequency-dependent flux shows $F_{\nu} \propto \nu^{1/3}$  behavior this can be done analytically (see e.g. Hryniewicz \& Czerny 2011), including the proportionality coefficient. However, in this work we calculate the monochromatic flux numerically, 
\begin{equation}
 L_{\nu} = \nu \int_{3 R_{Schw}}^{\infty} 4 \pi B_{\nu} (T_{eff}(r)) 2 \pi r dr,
\label{eq:mono}
\end{equation}
so no additional approximations are involved. Here $L_\nu$ represents the frequency times luminosity, as customarily used in the RM, and we adopt the viewing angle of  $60^{\circ}$ since the observed luminosity is usually determined assuming the isotropic emission, corresponding to $i = 60^{\circ}$.

Thus, for a fixed $T_{dust}$ and the range of black hole masses and accretion rates we calculate both the inner radius of the BLR and the monochromatic flux $L_{5100}$.

The applicability of the FRADO model as formulated above is based on several assumptions: the disk must be Keplerian, stationary, optically thick, and the total mass of the disk should not be much larger than the mass of the black hole. The first two  assumptions give us the knowledge of the angular momentum distribution independently from the disk structure, so the viscously generated flux is then independent from the viscosity mechanism. The third assumption guarantees the independence of the emited radiation on the disk structure - emission is uniquely determined by the local flux. However, the break down of any of these assumptions means that the model predictions are not reliable. The issue of too large total mass could be solved since the models based on the self-consistent description of the gravitational field in the presence of the massive disk have been studied (e.g. Hure 2002). However, the required computations are complex, so we use a spherically-symmetric potential and check the disk mass in exemplary solutions {\it a posteriori}.

\subsection{Self-gravity model}

The gravitational instability in the differentially rotating systems has been described by Safronov (1960) and Toomre (1964), and the criterion for the vertically averaged Keplerian disk is
\begin{equation}
Q = {c_s \Omega_K \over \pi G \Sigma} < 1,
\label{eq:q_glob}
\end{equation}
where $c_s$ is the sound speed, $\Omega_K$ is the local Keplerian angular velocity, and $\Sigma$ is the disk surface density. We can also use the criterion locally, within the disk, for example at the equatorial plane, and then this criterion can be expressed as
\begin{equation}
Q_{loc} = {M \over 2 \pi r^3 \rho_e} < 1,
\label{eq:q_loc}
\end{equation}
where $\rho_e$ is the local density in the disk at the equatorial plane.

In the outer parts of the disk this criterion is easily satisfied (see Rice 2016 for a recent review). The current studies indicate that in the accretion disks surrounding a supermassive black hole four characteristic regions likely appear:
\begin{itemize}
\item (i) inner region, were self-gravity is negligible and the disk structure is well described by the standard Shakura-Sunyaev disk (Shakura \& Sunyaev 1973)
\item (ii) second region, where the self-gravity leads to the development of the marginally self-gravitating disk ($Q \sim 1$), with gravity  working as a local viscosity law (Paczy\' nski 1978; Laughlin \& Rozyczka 1996; Collin \& Hure 1999; Lodato \& Rice 2004; Duschl \&  Britsch 2006;  Rafikov 2009)
\item (iii) third region where the self-gravity effects become global, the instability is violent leading to super-sonic turbulence; this region exists only at some evolutionary stages of an active nucleus (Kawakatu \& Wada 2008)
\item (iv) outer region, where self-gravity leads to star formation (Collin \& Zahn 1999, 2008; Thompson et al. 2005; Wutschik et al. 2013). 
\end{itemize}

In this paper we concentrate on the regions (i) and (ii), with particular attention payed to the transitions between (i) and (ii), and (ii) and (iii), as potentially related to the change in the disk structure leading to formation of the BLR.  

In order to determine the transition radii we need to know the disk structure, i.e. either the disk surface density or the density in the equatorial plane, depending on the criterion. Those quantities, unlike the disk effective temperature, are not set by the conservation laws for a stationary Keplerian disk but do depend on the assumptions about the viscosity, and on the details of the cooling efficiency. This last aspect, in turn, depends on the disk opacity which varies with the disk optical depth. In the inner parts the disk opacity is dominated by the electron scattering. However, in the outer parts other atomic processes play a role as well and they can affect both the disk surface density and the local density.

Thus, to determine the radius where the self-gravity effects start to be important we need to determine the disk structure, taking into account the proper description of the local opacity. Early papers considered the vertically-averaged disk structure (Paczy\' nski 1978, Rafikov 2009, Clarke 2009, Rice \& Armitage 2009). However, the predicted importance of the self-gravity depend considerably on the importance of the radiation pressure and the source of opacity. Simple criteria based on vertically-averaged solutions from Shakura \& Sunyaev (1973) imply different dependence of the transition radius on the black hole mass and accretion rate (see Appendix A), and provide very inaccurate approximation. 

We thus perform calculations of the disk vertical structure taking into account all the relevant processes, basically following the work of R\' o\. za\' nska et al. (1999). We neglect here the disk corona, but we introduce the self-gravity effects. Since we concentrate on the disk parts not too close to the black hole horizon, we neglect the advection term which in general should be also included (Abramowicz et al. 1988, S\c adowski et al. 2011).

\subsubsection{Local vertical structure of the accretion disk in local self-gravity region}

The local vertical structure of the classical accretion disk is set by three equations: vertical hydrostatic equilibrium, vertical energy transfer (radiative and eventually convective), vertical dissipation profile. The condition of the marginal self-gravity imposed a local condition of the density in the equatorial plane. However, as was shown in several papers (see Rice 2016 and the references therein), the dissipation in the marginally self-gravitating disk behaves to some extend as the modified  $\alpha$ viscosity. Therefore, both in the inner, non-selfgravity region and in the outer region we use the $\alpha$ prescription but in the inner region $\alpha$ is set as an arbitrary parameter while in the region (ii) $\alpha$ is determined by the condition that the local density in the equatorial plane satisfies the condition of the marginal self-gravity. Thus, effectively, $\alpha$ is constant is the innermost region, and rises in the region (ii).
  The transition to the region (iii) takes place when $\alpha$ reaches the value of 1 (Gammie 2001). Local equations are supplemented by the global conservation laws of the mass and angular momentum, and the radial pressure gradient are neglected so the circular motion velocity is determined by the local Keplerian orbits. 

We thus solve three equations describing the disk vertical structure at each radius, $r$. They represent the hydrostatic equilibrium, the energy transfer (radiative and convective), and the dissipation profile. The hydrostatic equilibrium is given by
\begin{equation}
{1 \over \rho} {dP \over dz} = - {GM z \over r^3} - 2 \pi G \Sigma_z
\label{eq:hydro}
\end{equation} 
Eq.~\ref{eq:hydro} differs from the corresponding formula in Shakura \& Sunyaev (1973) by the presence of second term on the right hand side representing the local self-gravity force,  compressing the disk and described as in Paczy\' nski (1978). The quantity $\Sigma_z$ is the surface density measured from the equatorial plane to the current value of the coordinate $z$, so it vanishes in the equatorial plane. We also include the effect of the disk mass being possibly comparable, or larger than the mass of the central black hole by simply adding the disk mass, within a given radius to the black hole mass. So here the mass $M$ includes both the mass of the black hole as well as the disk mass inside the radius $r$. More careful description of the global gravity field, including the departure from spherical symmetry, is beyond the scope of the current paper. The pressure, $P$, includes both the gas pressure and the radiation pressure, and the gas pressure calculation in the partially ionized zone includes the proper computation of the mean molecular weight. 

The energy transport in the vertical direction is described as
\begin{eqnarray}
F = F_{rad} = - {16 \sigma T^3 \over 3 \kappa \rho}, & &  ~~~~~ {\rm if } ~~~~ \nabla_{rad} \le \nabla_{ad}, \nonumber \\
F = F_{rad} + F_{conv}, & &  ~~~~~ {\rm if } ~~~~ \nabla_{rad} > \nabla_{ad}.
\label{eq:transfer}
\end{eqnarray}

We thus allow for convection to transport a fraction of the energy dissipated in the disk. This convection is described in a standard way, as in the stellar interior. 
The opacity $\kappa$ used here is the Rosseland mean, tabularized as a function of density and temperature. We use here the table from Seaton et al. (1994), obtained with atomic data from the Opacity Project, for log T $>$ 4.0, and tables from Alexander, Johnson \& Rypma (1983) for log T $<$ 3.8, and we interpolate between the two tables when necessary. Thus our opacity includes all the processes like electron scattering, free-free, atomic transitions for cosmically abundant elements (H, He, C, N, O, Ne, Na, Mg, Al, Si, S, A, Ca and Fe, in numerous ionization states), including the line blanketing, as well as the dust and grain opacity. Therefore, this opacity table can be applied in the inner as well as in the outer disk regions. We assumed solar metallicity. We neglect the magnetic energy transfer which in principle can be also important (e.g. Czerny et al. 2003, Begelman et al. 2015).

\begin{equation}
{dF \over dz} = {3 \over 2}\alpha P \Omega_K,
\label{eq:diss}
\end{equation}
where $\Omega_K$ is the local Keplerian angular velocity. Thus the energy production within the disk is described through the $\alpha$ parameter, as in the Shakura \& Sunyaev (1973).

In the innermost region the parameter $\alpha$ is set arbitrarily, in the region (ii) the required value of $\alpha$ is calculated from the condition that the disk is marginally self-gravitating, i.e. $\alpha$ becomes a function of the radius $r$. Numerical simulations (Forgan et al. 2011) justify this approach to the description of region (ii). The transition to the region (iii) happens when $\alpha$ becomes equal 1. Simple continuous disk description does not apply beyond this radius, where clumps form due to the efficient cooling.

\subsubsection{Global conservations laws and boundary conditions for a vertical structure computing}

The integration of the disk vertical structure is performed from the top downward, to the disk equatorial plane. The standard conservation laws of mass and angular momentum form the the boundary conditions, as in Shakura-Sunyaev disk. The total dissipated flux at the radius $r$ is given by
\begin{equation}
F_{tot} = {3 G M \dot M \over 8 \pi r^3} \bigl(1 - \sqrt{3 R_{Schw} \over r}\bigr),
\label{eq:flux2}
\end{equation}
and the effective temperature is thus given by
\begin{equation}
T_{eff} = ({F_{tot} \over \sigma_B})^{1/4},
\label{eq:temp2}
\end{equation}
where $\sigma_B$ is the Stefan-Boltzman constant. Since we use the diffusion approximation for the energy transfer (see Eq.~\ref{eq:transfer}) the temperature at the outer edge of the disk atmosphere, $T_0$, is lower by a factor $2^{1/4}$ than the effective temperature, as in standard Eddington approximation (Mihalas 1978). The effective temperature is achieved at the optical depth $\tau = 2/3$. The density at the disk surface is assumed to be $10^{-16}$ g cm$^{-3}$ instead of 0 to allow for the opacity computation. The disk thickness is determined iteratively from the condition that the flux dissipated in the disk and described by Eq.~\ref{eq:transfer} reaches zero at the equatorial plane. The integration is
performed by the second-order Runge-Kutta scheme with adaptive step size.

In the region of the marginal self-gravity additional iteration is necessary, that is the adjustment of the viscosity parameter $\alpha$, then dependent on the disk radius, to ensure that the disk density in the equatorial plane satisfies the condition of the marginal self-gravity $Q_{loc} = 1$ (see Eq.~\ref{eq:q_loc}). Our approach is basically equivalent to the method used by Rafikov (2009), Clarke (2009) and Rice \& Armitage (2009) but these authors did not consider the vertical structure of the disk. 

In the case of self-gravity model, we also calculate the monochromatic luminosity at 5100 \AA~ using Eq.~\ref{eq:mono}.

Application of Eqs.~\ref{eq:flux2} and \ref{eq:temp2} to the disk with self-gravity means that the prediction of the FRADO model for such disk is unchanged by the local self-gravity effect although the internal disk structure may be strongly affected. Introduced assumption of the local black body emission has to be checked a posteriori since self-gravity may strongly reduce the disk surface density. 

\subsection{Observational data}

We use the time delay measurements of H$\beta$ emission line summarized in
Du
et al. (2015, 2016), which are compiled from the reverberation campaigns
performed by different groups (Barth et al. 2013, Bentz et al. 2006, 2007,
2009b, 2014, Collier et al. 1998, Denney et al.
2006, 2010, Dietrich et al. 1993, 1998, 2012, Du et al. 2014, 2015, 2016,
Grier
et al.  2012, Kaspi et al. 2000, Korista et al. 1995,
Pei et al. 2014, Peterson et al.  1991, 1992, 1994, 1998, 1999, 2002, 2014,
Santos-Lle\'{o} et al. 1997, 2001, Stirpe et al. 1994, Wang et al. 2014). 
The time delays represent the emissivity-weighted radii of the BLR in those RM objects.
The contributions of host galaxies to their 5100 \AA\
luminosities
have been subtracted mainly using high-resolution images from {\it Hubble
Space
Telescope} (Bentz et al. 2009a, 2013, Du et al. 2014, Wang et al.  2014), or
adopting the empirical relationship for several objects (see Du et al. 2015,
2016). The cosmological parameters used are: Hubble constant $H_0 = 67$ km
s$^{-1}$ Mpc$^{-1}$, $\Omega_{\Lambda} = 0.68$, and $\Omega_M = 0.32$.

\section{Results}

We consider two likely models of the BLR formation with the aim to compare the predicted location of the BLR with the time delays between the continuum and the BLR emission lines, and to answer which of the two models better represent the observational data coming so far from the reverberation monitoring.

\subsection{The dust appearance in the disk atmosphere: FRADO model}

\begin{figure}
    \centering
 \includegraphics[width=0.95\hsize]{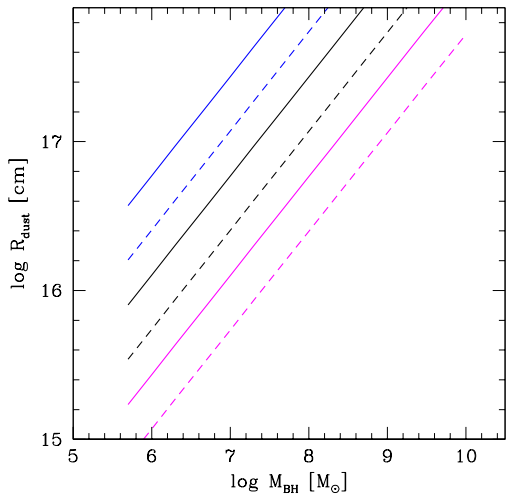}
    \caption{The dust sublimation radius  as a function of the black hole mass, 
for three values of the accretion rate in dimensionless units (magenta, black and blue
 for $\dot m = 0.01, 1$ and 100, correspondingly, and two value of the dust sublimation temperature: 800 K (continuous line) and 1500 K (dashed line).}
    \label{fig:rdust}
\end{figure}

We perform set of computations for a range of black hole masses and accretion rates with the aim to locate the radii where the effective temperature is equal to 800 K and 1500 K which we consider as a lower and upper limit for the dust sublimation temperature. The black hole masses are taken from the range $3 \times 10^5 M_{\odot}$ to $10^{10} M_{\odot}$. We use the dimensionless accretion rate
\begin{equation}
\dot m = {\dot M \over \dot M_{Edd}},~~~~ \dot M_{Edd} = 1.27 \times 10^{18} \bigl({ M \over M_{\odot}}\bigr)~~~~ {\rm [g~s}^{-1}].
\end{equation}
and we do the computations for $\dot m = 0.01$, 1.0, and 100. Quasars and bright AGN are usually though to be within 0.01 - 1.0 range (e.g. Kelly et al. 2010; Suh et al. 2015). However, there are indications that a fraction of sources may be at highly super-Eddington accretion rates (Collin et al. 2002; Abolmasov \& Shakura 2012; Du et al. 2014, 2015, 2016; Wang et al. 2014). The disk model we use is not well adjusted to highly super-Eddington accretion rates so the results for the last value should be treated with much reservation, but we show then to indicate the possible trend.

The predicted location of the inner radius of the BLR within the frame of the FRADO model is given in Fig.~\ref{fig:rdust}. The location depends strongly on the accretion rate $\dot m$ as well as on the adopted dust temperature.

We now assume that the BLR size can be simply interpreted as the time delay, without any geometrical factors connected with the source inclination and the extension of the BLR. We also assume that the measured monochromatic luminosity represents the mean isotropic radiation flux. Under such assumptions we obtain an expected delay versus the monochromatic source luminosity at 5100 \AA, and we can compare it to the measured time delays in the sample of AGN.

\begin{figure}
    \centering
 \includegraphics[width=0.95\hsize]{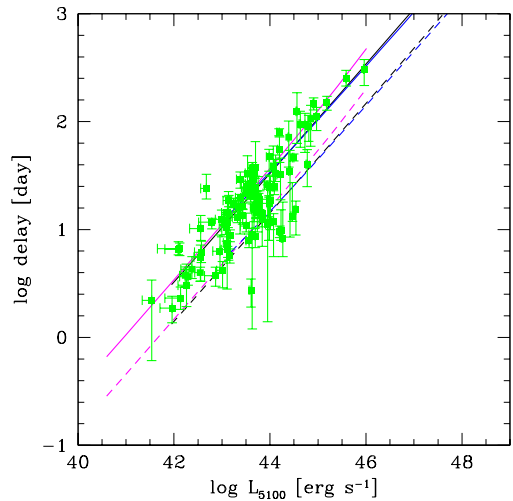}
    \caption{The expected time delay from FRADO model as a function of the monochromatic luminosity at 5100 \AA,
for three values of the accretion rate in dimensionless units (magenta, black and blue
 for $\dot m = 0.01,$ 1 and 100, correspondingly, and two value of the dust sublimation temperature: 800 K (continuous line) and 
1500 K (dashed line). Green points are observational data from Du et al. (2016).}
    \label{fig:del_dust}
\end{figure}
 
The result is shown in Fig.~\ref{fig:del_dust}. The picture looks different now, when the horizontal axis is replaced with the monochromatic luminosity. In this case the dependence on the accretion rate practically disappears. There is a slight departure from the strait line in the case of large black hole mass and low accretion rate since in this case the disk is colder, and the emission at 5100 \AA ~ is relatively close to the peak of the disk spectrum instead of following the classical Shakura-Sunyaev behavior at longer wavelengths, $F_{\nu} \propto \nu^{1/3}$. However, when no such departure is seen, the delay - luminosity relation follows exactly the pattern $\tau_{del} \propto L_{5100}$, as outlined in Czerny \& Hryniewicz (2011). The predictions for the highest accretion rate, $\dot m$, are strongly biased: the position of the sublimation radius is well represented but the monochromatic luminosity is overpredicted due to the negligence of the advection, important in the innermost part of the disk. This bias should be lower than a factor 10 since the total luminosity of a slim disk saturates at $\sim 10 L_{Edd}$.

The comparison of the model prediction with the set of the measuerd time delays looks favorable for the FRADO model (see Fig~\ref{fig:del_dust}). Intermediate temperature between the adopted 800 K and 1500 K seems appropriate. There are a few outliers but overall most of the sources follow the trend. The best fit is indeed obtained for the dust temperature $T_{dust} = 899.6$ K, not shown on the plot.  
%(best_temperature.f in progress)

Obtained dust temperature depends on the cosmological model used to determine the observed $L_{5100}$ luminosities. In general, it also depends on the viewing angle, $i$, here assumed to be $60^{\circ}$ and the exact geometry of the BLR. The inclination enters into the disk flux through a $\cos (i)$ for a geometrically flat disk, as well as in the time delay if the cloud distribution has a 3-D shape and is spatially extended. Here no such effects were introduced, and the simple relation between the inner BLR radius and the time delay was adopted. The reliable determination of the inclinations in radio-quiet AGN still poses a considerable problem (see e.g. Middleton et al. 2016, Marin 2016). If the typical inclination of the source is 30$^\circ$ instead of  $60^{\circ}$, this will result is a systematic offset in the luminosity by a factor of $2 \cos 30^\circ \sim 1.732$ provided by the model but the same offset will be included in the derivation of the luminosity from the data.  This systematic shift of the model and the data shift would lead to a different value (1100 K) for the dust sublimation temperature, best representing the data. The observed dispersion around the model lines in Fig.~\ref{fig:del_dust} is, in principle, due to the combination of the possible differences in the intrinsic geometry, viewing angles and dust sublimation temperature.  The dispersion around the relation is only 0.25 dex, and the intrinsic scatter is only 0.21 (Du et al. 2016) which can be fully accounted for by the dispersion of the viewing angles, so additional intrinsic dispersion in the dust sublimation temperature seems unlikely.

\subsection{The radius of the self-gravity onset}

\begin{figure}
    \centering
 \includegraphics[width=0.95\hsize]{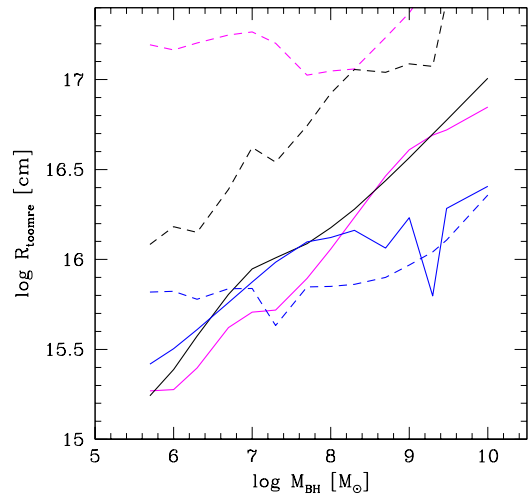}
    \caption{The transition radius between the zone (i) and (ii), i.e. radius where the Toomre parameter becomes smaller than 1, as a function of the black hole mass (continuous lines), for three values of the accretion rate in dimensionless units (magenta, black and blue continuous lines for $\dot m = 0.01, 1$ and 100, correspondingly, and the viscosity parameter $\alpha = 0.01$. Dashed lines mark the radii where the disk mass becomes larger than the central mass of the black hole. }
    \label{fig:r_sg_glob}
\end{figure}

The first effects of the self gravity may appear when the Toomre parameter given by Eq.~\ref{eq:q_loc} becomes smaller than 1. The position of this radius, calculated from the disk model without self-gravity effects included is shown in Fig.~\ref{fig:r_sg_glob}. The transition radius between the zone (i) and zone (ii) only weakly depends on the Eddington ratio at sub-Eddinton to Eddington luminosities, and rises with the black hole mass, if measured in absolute units. The rise, however, is much slower than linear, so the value of the transition radius measured in Schwarzschild units drops with mass from $10^4$ for $5 \times 10^5 M_{\odot}$ down to 10 for $10^{10} M_{\odot}$. When the accretion rate becomes much higher, the transition radius seems to level off for larger masses, and the whole disk becomes self-gravitating for $10^{10} M_{\odot}$. This thus happens below the absolute limit of $5 \times 10^{10} M_{\odot} $ for the mass of the accreting black hole derived by King (2016) at the basis of very approximate formulae for self-gravity effects. 

The difference between the local and global criterion (see Eq.~\ref{eq:q_loc} and \ref{eq:q_glob}) is small; usually the local instability close to the equatorial plane appears first, but it is followed by the whole disk instability at a distance not larger than by a factor 2. 

The radius, where the total disk mass inside that radius is equal to the black hole mass is in general larger than the transition between (i) and (ii) region. However, for the extreme case of very large accretion rate ($\dot m = 100$) the two radii are comparable (see dashed lines Fig.~\ref{fig:r_sg_glob}). Our solution may not be accurate in the region where the disk mass inside the studied radius becomes larger than the black hole mass. Here, more carefull approach, following for example Hure (2002) would provide better approximation.

\begin{figure}
    \centering
 \includegraphics[width=0.95\hsize]{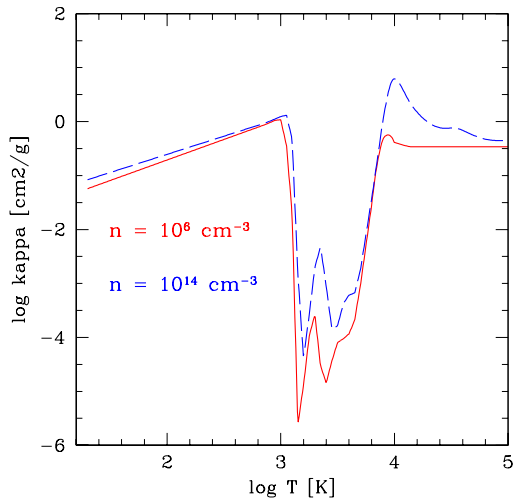}
    \caption{The dependence of  the Rosseland mean opacity used in our computations on the local temperature, $T$, for two value od the density $10^6$ cm$^{-3}$ and  $10^{14}$ cm$^{-3}$. The drop in the opacity just above $\sim 1000$ K is caused by the gas becoming neutral (no electron scattering) while the dust and grain opacities sets in at still lower temperature. }
    \label{fig:opacity_general}
\end{figure}

The dependence of the disk properties on the radius is not as smooth as in the case of the disk effective temperature in FRADO model. The local disk structure depends on the opacity, which in turn has a complex dependence on the local density and temperature which vary both in the vertical and in the radial direction. The opacities used by us are reliable and show overall similarity to the opacities used by Thompson et al (2005), as illustrated in Fig.~\ref{fig:opacity_general}.

\begin{figure}
    \centering
\includegraphics[width=1.5\hsize]{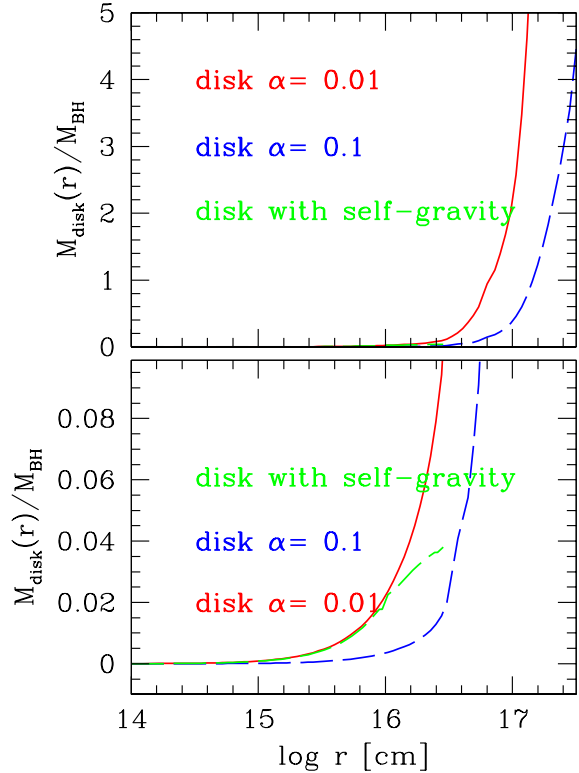}
    \caption{The disk mass inside the radius $r$ as a function of the radius, for a disk without self-gravity effects and two values of the viscosity parameter $\alpha$ (0.01 - red line, 0.1 - blue line), and the disk with self-gravity effects and $\alpha = 0.01$ in the innermost part (green line, visible only in the expanded lower panel), for the black hole mass $M = 3 \times 10^{7} M_{\odot}$ and accretion rate $\dot m = 1.0$.}
    \label{fig:masa_calkowana}
\end{figure}

The results depend on the adopted value of the viscosity coefficient in the innermost region. We assumed $\alpha = 0.01$ as our reference value. Higher value of the viscosity parameter leads to lower value of the disk surface density, the total disk mass decreases at a given radius, and the radius where the disk mass equals the black hole mass moves outward (see Fig.~\ref{fig:masa_calkowana}). A change in $\alpha$ from 0.01 to 0.1 moves the transition radius between the region(i) and (ii) by a factor of 2 for typical parameter values. 

\begin{figure}
    \centering
 \includegraphics[width=0.95\hsize]{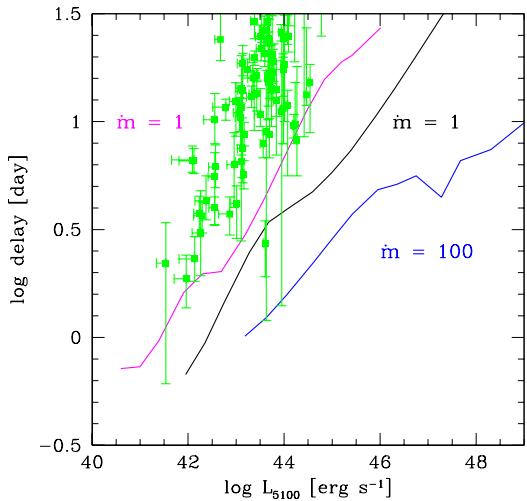}
    \caption{The transition radius measured in light days between the zone (i) and (ii), where the Toomre parameter becomes smaller than 1, as a function of the monochromatic disk flux at 5100 \AA, for three values of the accretion rate in dimensionless units ($\dot m = 0.01$ - magenta, 1.0 - black, and 100 - blue). Green points are observational data from Du et al. 2016.}
    \label{fig:Toomre}
\end{figure}

From the observational point of view, it is more convenient to measure this radius in light days, and in reference to the disk monochromatic flux at 5100 \AA, as frequently used in reverberation campaign (see Fig.~\ref{fig:Toomre}), as we did for FRADO model. The time delay predicted by a transition between the region (i) and region (ii) is however much shorter than previously, from 1 to 30 days even for high luminosity sources. This transition is thus not connected with the BLR formation which happens further out.

The second important radius is marked by the total disk mass becoming equal to the black hole mass. This radius is located further out (see Fig.~\ref{fig:r_sg_glob}). However, here the radius where the total mass equals the central mass have been computed without the self-gravity effects, i.e. not self-consistently since this radius is located beyond the region (ii).

The transition between the region (ii) and region (iii) happens further down, at larger radii. We thus calculated several disk models, with self-gravity included, up to the radius where $\alpha = 1$, i.e. the self-gravitating disk stops cooling efficiently to prevent the fragmentation. Here the self-gravity effects are very important and they modify the disk structure considerably, in comparison with the standard model without the self-gravity effects. 

\begin{figure}
    \centering
 \includegraphics[width=1.8\hsize]{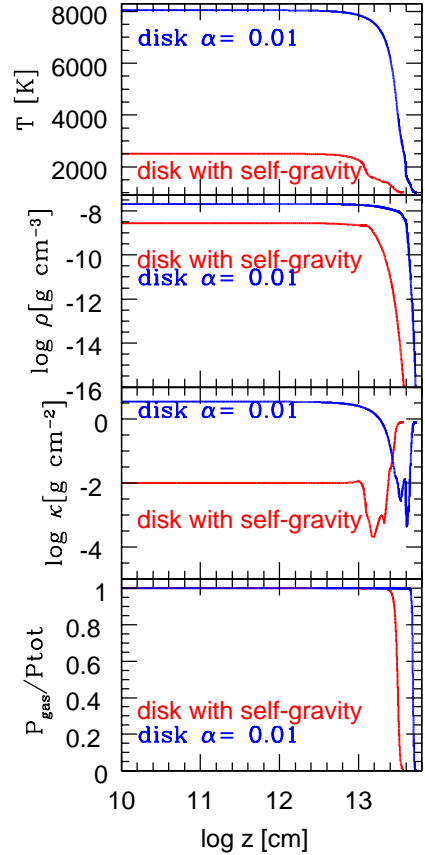}
    \caption{The vertical structure of the accretion disk close to the outer edge of the self-gravitation region (ii) in comparison to the model without self-gravity effects: temperature, $T$, density, $\rho$,  opacity, $\kappa$, and the gas to the total pressure ratio plotted as a function of $z$ measuring the distance from the equatorial plane. $M = 3 \times 10^7 M_{\odot}$, $\dot m = 0.01$, $R = 1661 R_{Schw}$.}
    \label{fig:disk_structure}
\end{figure}

In Fig.~\ref{fig:disk_structure} we show a comparison of the disk vertical structure close to the border between region (ii) and region (iii) for an exemplary black hole mass and accretion rate, and we compare it to the standard disk structure calculated without any self-gravity effects. This specific solution has the temperature at the surface 1030 K (effective temperature 1225 K). The self-gravitating disk is then almost isothermal, with equatorial temperature only two times higher than the surface temperature while the standard disk is much hotter inside, denser, the surface density is also higher but the geometrical thickness of the disk is very similar in both solutions. The  Rosseland mean opacity is very much different in these two solutions since the standard disk is hot and fully ionized, with opacity provided by numerous bound-free and line transitions while most of the self-gravitating disk is in the region of the opacity drop (see Fig.~\ref{fig:opacity_general}). The total optical depth of the self-gravitating disk is more than three orders of magnitude lower but the disk is not optically thin: the total optical depth is $\sim 300$. Both solutions are gas-dominated apart from the uppermost region where the density is very low.

This decrease in the disk surface density in self-gravitating solution in comparison with the standard disk is responsible for the slower rise in the total disk mass. We show that trend in Fig.~\ref{fig:masa_calkowana} marked with a green dashed line. The self-gravitating solution is not extended beyond the transition radius from the region (ii) to (iii) while the standard disk was formally extended beyond that region. The curve is only visible in the lower, expanded part of the plot as for the adopted parameters the transition to the region (iii) happens at $\log R = 16.42$ cm. There, the rise in the total mass is not yet considerable but the assumptions of the model used in this paper break down beyond that radius. This is true for all the solutions with accretion rate $\dot m \leq 1$: the transition to the region (iii) happens (usually well) before the disk mass becomes comparable to the black hole mass. For the most extreme case, $\dot m = 100$, and for the black hole mass of $10^7 M_{\odot}$, or larger, the total mass becomes larger than the black hole mass at smaller radii than the local self-gravity criteria (Eqs.~\ref{eq:q_glob} and \ref{eq:q_loc}) are satisfied. Much more careful treatment of the disk accretion flow is certainly necessary in this parameter space.

\begin{figure}
    \centering
\includegraphics[width=1.5\hsize]{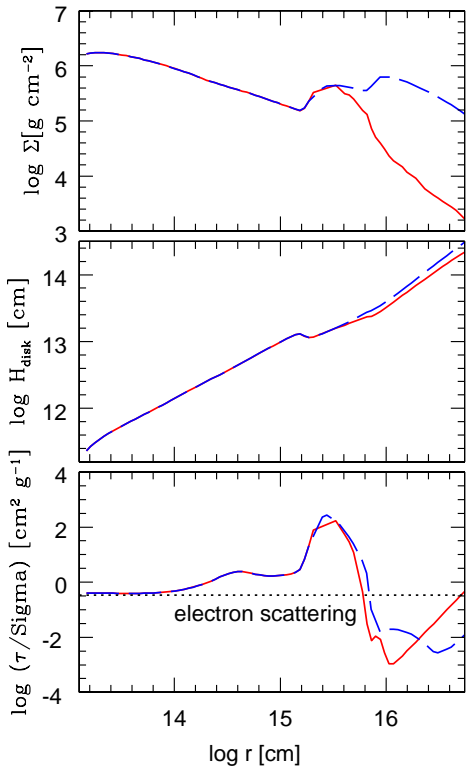}
    \caption{The exemplary radial dependence of the disk surface density, thickness and the effective opacity for a disk $M = 10^7 M_{\odot}$, $\dot m = 0.01$ (red continuous line), in comparison to a standard  $\alpha = 0.01$ disk (blue long dashed line); the self-gravity effects become essential ($\alpha$ rises) from log r = 15.5 cm.}
    \label{fig:radial}
\end{figure}

An example how the self-gravity affects the disk at various distances from the black hole is shown in Fig.~\ref{fig:radial}. Both disks do not practically differ in the inner region although two self-gravity effects were included at all radii: the rise of the central mass with the radius and the vertical component due to self-gravity. However, essential differences arise only after crossing the border between (i) and (ii) region, with the rise of the effective $\alpha$. The disk thickness does not decrease significantly despite the drop in the equatorial pressure in comparison with the standard disk. This is the effect noticed by Thompson et al. (2005), although their simplified description has lead to the disk height to the radius ratio of order of 100, thus implying a narrow funnel formed by the dusty disk around the symmetry axis. Our method of solving for the disk vertical structure implies moderate and realistic disk height to the radius ratio of order of a few per cent although the disk atmosphere is already dust-dominated while the disk interior is in the opacity-dip zone.

\begin{figure}
    \centering
 \includegraphics[width=0.95\hsize]{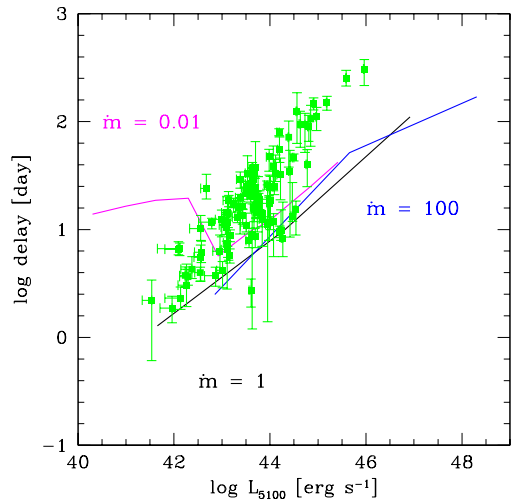}
    \caption{The transition radius between the region (ii) and (iii)  measured in light days, as a function of the monochromatic disk flux at 5100 \AA, for three values of the accretion rate in dimensionless units (0.01 - magenta, 1.0 - black, 100 - blue). Green points from Du et al. (2016)}
    \label{fig:del_sg_regionII}
\end{figure}

Fig.~\ref{fig:del_sg_regionII} shows the expected delay if the BLR forms where the self-gravity becomes more violent, i.e. at the transition from the region (ii) to region (iii). The delay is a smooth and monotonic function for higher accretion rates but for the low accretion rate the dependence on  $L_{5100}$ is more complex due to the opacity, as with the decrease of the temperature the opacity first shows a rise at intermediate temperatures (this is sometimes call a metallicity bump), and then a rapid decrease, later inverted when dust and grains set in (see Fig.~\ref{fig:radial}). For higher accretion rates the transition radius happens always at a radius with the effective temperature higher than $1000 K$, i.e. in the region where there is no dust in the disk atmosphere. However, for $\dot m = 0.01$ the relative position of the dust sublimation radius and the transition (ii) to (iii) radius depend on the black hole mass. For low black hole mass the dust sets in at smaller radii, and the effective temperature at the transition radius is low. For example, for the black hole mass $3 \times 10^5 M_{\odot}$ the transition between (ii) and (iii) region happens at $4.1 \times 10^5 R_{Schw}$, where the effective temperature is extremely low, only 64 K. This temperature rises with the black hole mass to 285 K for $10^7 M_{\odot}$, and to 1000 K at $3 \times 10^7 M_{\odot}$. 

Since the transition between the regions (ii) and (iii) takes place at a larger radius than the transition between (i) and (ii) we checked if the model is still self-consistent. The disk is always optically thick in the whole studied region. The disk mass does not rise too strongly in the marginal self-gravity solution (see Fig.~\ref{fig:radial}). The total mass of the disk up to the zone (iii) boundary, however, becomes quite large in the case of very large accretion rates and high masses. For $\dot m = 1$ the disk mass reaches 50\% of the black hole mass in case of high mass black hole $M = 3 \times 10^9 M_{\odot}$. This still does not pose much of the problem to the model: disk mass in included (using a spherically symmetric approximation), and the local departure from the Keplerian motion is at the level of 0.5 \%, or less. However, for $\dot m = 100$ and $ M = 3 \times 10^9 M_{\odot}$, the disk total mass exceeds the black hole mass by a factor 20, the relative correction to the Keplerian motion due to the radial pressure gradient is 70\%, and the heat advection carries 25\% of the energy. Thus, if such extreme accretion rates in quasars containing very massive black holes are confirmed, the model has to be refined following the use of the slim disk (Abramowicz et al. 1988) concept combined with a better description of the gravity (Hure 2002). Currently, most quasars are considered to be radiating between 0.01 and 1 in Eddington unit (see e.g. Kelly et al. 2010).

We can again compare the predicted delay, this time based on the assumption that the transition between the region (ii) and (iii) is responsible for the onset of the turbulence and subsequent formation of the BLR with the data sample. The data points are not consistent with expectations. The expected time delays are too short in comparison with the model for luminous objects. At lower luminosity end, the agreement seems better. In that region the sources having higher mass (lower $\dot m$) for the same $L_{5100}$ should show longer time delay than smaller mass objects.

\section{Discussion}

We calculated the expected time delays between the disk continuum determined at 5100 \AA~ and the line emission coming from the BLR in two scenarios: FRADO model and self-gravity model. Both scenarios predict an enhanced turbulence/outflow in the outer part of the disk but the mechanism, and the exact location are different. In FRADO model this happens in the region where the dust can be present in the disk atmosphere, and in the self-gravity model it happens where the clumpiness develops in the disk due to self-gravity effects. We calculated the disk structure in 1-D approximation, i.e. the disk vertical structure has been calculated assuming a geometrically thin stationary Keplerian disk. We used the predicted radius as a measure of the time delay, without specific geometrical factors due to the viewing angle, or extension of the BLR. 

The predictions of the FRADO model do not depend on the disk structure since the effective temperature in the stationary Keplerian disk is determined by the accretion rate and black hole mass directly by the conservation laws. The predictions of the self-gravity scenario depend on the disk description. However, we use the best possible approximation to the 1-D disk structure by carefully including the radiation and gas pressure, and in particular using the opacities which include all important atomic transitions, opacity due to the molecules, and the dust grains. 

In the case of FRADO model, the expected delay does not depend on the accretion rate and the mass separately, just on the monochromatic luminosity at 5100 \AA~and basically follows the power law behavior, $R_{BLR} \propto L_{5100}^{1/2}$. The proportionality constant depends on the assumed dust sublimation temperature, and the value of 900 K for the maximum dust temperature well represents the time delays measured in the AGN sample. The comparison between the data and the model is thus favorable for the FRADO scenario.

In the current paper we neglected the issue of the viewing angle with respect to the symmetry axis in various AGN.  In our computations this viewing angle, $i$, was fixed at 60 deg since usually the luminosity is determined assuming the isotropic emission. However, since the disk emission is proportional to the $\cos (i)$, and the time delay is thus proportional to $\cos (i)^{-1/2}$. If the actual distribution of the viewing angles in type 1 AGN is between 0 ad 45 deg, this factor could account for the dispersion of order of 14 \%, i.e. 0.06 deg in the log space. The dependence of the dust temperature is stronger, the delay is proportional to $T_{dust}^{-4/3}$. Thus a significant dispersion in the dust sublimation temperature between various objects could easily lean to much larger dispersion in the delay-luminosity relation than observed. It supports the view that the dust chemical composition in all AGN is universal. 

In the case of the self-gravity model, the predicted delay depends not just on the monochromatic luminosity but also on the accretion rate, particularly for small black hole masses. The slope if the predicted delay-luminosity relation is more shallow, and in general, the measurements of the time delay are not in agreement with the self-gravity scenario.

Our calculations have been performed using a model of the Keplerian disk. If the disk becomes massive in its innermost part the motion of the plasma is not Keplerian, as recognized by Abramowicz et al. (1984)  and reviewed by Karas et al. (2004). The expression used by us in Eq.~\ref{eq:hydro} assumes local plane-paralel approximation but this is a reasonable approximation unless the disk is very centrally  condensed (Trova et al. 2014). In typical stationary AGN disk the Keplerian approximation seems a reasonable approach. Within this frame, using the FRADO scenario we thus have a satisfactory explanation of the position of the inner edge of the BLR as related to the sublimation radius.

The problem remains, however, that the self-gravity radius determined from the model is significantly smaller that the BLR for high black hole masses, while larger than the BLR inner region for small masses. Still, we we not seem to notice a significant difference between the BLR properties for small and large black hole masses at the same accretion rate. 

Numerical simulations of the time-dependent behavior of self-gravitation accretion disk show that the region (ii) is still overall similar to the standard disk although some spiral wave patters may develop (Lodato \& Rice 2005). However, the region (iii) is predicted to be more violent. This region formally might correspond to $\alpha > 1$. The cooling is fast there, and the relative perturbations of the disk mean density ($\delta \Sigma /\Sigma \approx {\sqrt \alpha}$, Rice et al. 2011) are large, and the disk should become clumpy. If the existence of the BLR is tightly related to the presence of the underlying cold accretion  disk, as frequently argued (e.g. Czerny et al. 2004, Balmaverde \& Capetti 2014, and the references therein), the significant change in the structure of the underlying disk should affect somehow the BLR. However, if the BLR lines form directly in the inflowing material (see Gaskell \& Goosmann 2016) or come from the material temporarily ejected from the disk by passing stars (Zurek, Siemiginowska \& Colgate 1994) then the physical state of the underlying disk becomes relatively unimportant. However, in this case we do not have any explanation for the observed correlation between the BLR size and the monochromatic flux.

\acknowledgments
{\small 
The project was partially supported by the Polish Funding Agency National Science Centre, project 2015/17/B/ST9/03436/ (OPUS 9). 
This work was supported by the Strategic Priority Research Program—The Emergence of
Cosmological Structures of the Chinese Academy of Sciences, Grant No.
XDB09000000grant 2016, by NSFC grants NSFC-11233003, -11503026, and YFA0400702 from
the Ministry of Science and Technology of China.
%and by  European Union Seventh Framework Program (FP7/2007-2013)  under the grant agreement No. 312789.
}

%\clearpage

%%\appendix

\section*{Appendix A: Self-gravity region in the Shakura-Sunyaev disk}

The famous analytical solution to the disk structure of Shakura \& Sunyaev (1974) allows us to derive the location of the self-gravity onset
also analytically. The problem is that those analytic solutions form three regions, depending on the role of the radiation pressure and the
dominant character of opacity. Assuming the Toomre local criterion (Eq.~\ref{eq:q_loc}) in each of these regions we obtain the radius for the onset of the instability expressed in the Shakura-Sunyaev units: 
\begin{equation}
 \mathcal{R}=1.17\times10^{4}~\alpha^{2/9}~\dot{\mathcal{M}}^{4/9}~m_7^{-2/9}, ~~~~{\rm (region~~a)}
\label{eq:anal}
\end{equation}

\begin{equation}
 \mathcal{R}=3.09\times10^{4}~\alpha^{14/27}~\dot{\mathcal{M}}^{-8/27}~m_7^{-26/27},
~~~~{\rm (region~~b)}
\end{equation}

\begin{equation}
\mathcal{R} =2.69\times10^{3}~\alpha^{28/45}~\dot{\mathcal{M}}^{-22/45}~m_7^{-52/45},
~~~~{\rm (region~~c)}
\end{equation}
where $\mathcal{R} = R/(3 R_{Schw})$, $m_7 = M/(10^7 M_{\odot}$), $\dot{\mathcal{M}} = {\dot M} /\dot M_{crit}$, and  $\dot M_{crit} = 3 \times 10^{-8} M_{\odot} yr^{-1} \times (M/M_{\odot})$, as used in the original paper. Formula (\ref{eq:anal})  is frequently used for the general purpose (e.g. Laor \& Netzer 1989; 
Netzer 2015). 

\begin{figure}
    \centering
\includegraphics[width=0.95\hsize]{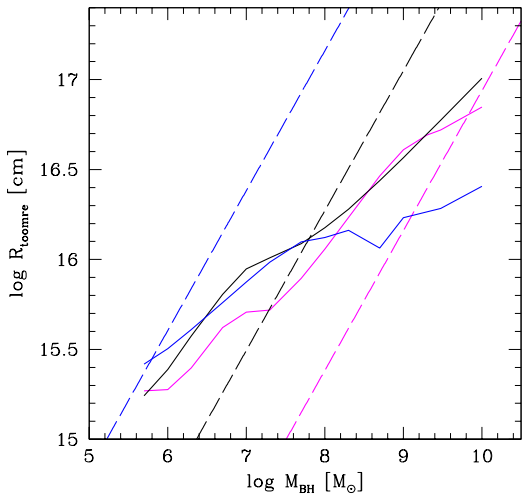}
    \caption{The comparison of the numerical results (continuous line) for the location of the self-gravity 
zone with analytical approximation (dashed line) for three values of the accretion rate, 0.01 (magenta), 1 (black) and 100 (blue) using region (a) approximation. Expressions for region (b) and (c) predict the self-gravity as a very weak or decreasing function of the black hole mass which does not match better the numerical results. }
    \label{fig:appendix}
\end{figure}

Since time delay measurement refers to the absolute units, and we use slightly different definition of the dimensional accretion rate throughout the paper, we can express those values conveniently as:
\begin{equation}
R_{SG} = 1.03\times10^{16}~\alpha^{2/9}~\dot{\mathcal{M}}^{4/9}~m_7^{7/9}~{\rm [cm]}, {\rm (region~~a)}
\label{eq:anal_cm}
\end{equation}
\begin{equation}
R_{SG} = 2.72\times10^{16}~\alpha^{14/27}~\dot{\mathcal{M}}^{-8/27}~m_7^{1/27}~{\rm [cm]}, {\rm (region~~b)}
\end{equation}
\begin{equation}
R_{SG} = 2.38\times10^{16}~\alpha^{28/45}~\dot{\mathcal{M}}^{-22/45}~m_7^{-7/45}~{\rm [cm]},  {\rm (region~~c)}
\end{equation}
but none of those approximations is appropriate in general. Equations above also describe the transition between region (ii) and region (iii) when supplemented with condition $\alpha = 1$. In this region the radiation pressure and the gas pressure are comparable, and both atomic transitions and electron scattering contribute to the opacity. The comparison between the analytic expression \ref{eq:anal_cm} and the numerical values is shown in Fig.~\ref{fig:appendix} where we used the same units for $\dot m$ as in the body of the paper (i.e. $\dot M_{Edd} = 1.27 \times 10^{18}$ g s$^{-1}$) which is slightly different from the critical accretion rate in the original paper of Shakura \& Sunyaev (1973), $\dot m_{crit} = 1.89 \times 10^{18}$ g s$^{-1}$.

More advanced but still analytical formulae for the disk structure were given for example by Vaidya et al. (2009) but this required the division of the disk into even larger number of zones than in Shakura \& Sunyaev (1973).

%{\bf [De Liang,Wu derived it]}

\bibliographystyle{apj}
\bibliography{refs}
\end{document}